\documentclass[final]{svjour2}
\usepackage{graphicx}
\usepackage{epstopdf}
\usepackage{amssymb,bm}
\usepackage{mathptmx}
\usepackage[numbers]{natbib}
\makeatletter
\journalname{Journal of Low Temperature Physics}
\bibpunct{}{}{,}{s}{}{,}

\begin{document}

\newcommand{\hdblarrow}{H\makebox[0.9ex][l]{$\downdownarrows$}-}
\title{Vortices and Dynamics in Trapped Bose-Einstein Condensates}

\author{Alexander L.\  Fetter}

\institute{GLAM, McCullough Building\\
Stanford, CA 94305-4045\\
\email{fetter@stanford.edu}}
%\footnote{Article belongs to Special issue SUR2010 Ð Section III}

%\date{28.04.2010}

\maketitle


\begin{abstract}

I review the basic physics of ultracold dilute trapped atomic gases, with emphasis on Bose-Einstein condensation and quantized vortices.  The hydrodynamic form of the Gross-Pitaevskii equation (a nonlinear Schr{\"o}dinger equation) illuminates the role of the density and the quantum-mechanical phase.  One unique feature of these experimental systems is the opportunity to study the dynamics of vortices in real time, in contrast to typical experiments on superfluid $^4$He. I discuss three specific examples (precession of single vortices, motion of vortex dipoles, and Tkachenko oscillations of a vortex array).  Other unusual features include the study of quantum turbulence and the behavior for rapid rotation, when the  vortices form dense regular arrays.  Ultimately, the system is predicted to make a quantum phase transition to various  highly correlated many-body states (analogous to bosonic quantum Hall states) that are not  superfluid and do not have  condensate wave functions.  At present, this transition remains elusive.  Conceivably, laser-induced synthetic vector potentials can serve to reach this intriguing phase transition.

\end{abstract}

\section{Introduction}
Consider a uniform gas of particles with mass $M$ and number density $n$.  The interparticle spacing is $\sim n^{-1/3}$, and there are two different approaches to discuss the onset of quantum degeneracy (for general background, see \cite{Dalf99,Fett01a,Fett02,Peth08,Fett09}).

Start with an atomic-physics perspective:  the mean thermal energy $p^2/2M\approx k_B T$ yields a mean thermal momentum $p\sim\sqrt{Mk_B T}$, and the familiar de Broglie relation $\lambda\sim h/p$ gives the mean thermal wavelength $\lambda\sim \hbar/\sqrt{Mk_BT}$.  Compare $\lambda$ with the interparticle spacing $n^{-1/3}$.  In the classical limit (high temperature, short wavelength),  we have $\lambda \ll n^{-1/3}$.  Hence quantum diffraction is negligible, similar to ray optics for light when diffraction is unimportant.  It is convenient to define the dimensionless parameter $n\lambda^3$, known as the ``phase-space density;''  this parameter is small in the classical limit since $\lambda \to 0$ when $\hbar\to 0$ or $T\to \infty$.  As the temperature falls at fixed $n$, the thermal wavelength grows, and quantum degeneracy appears when $n\lambda^3$ is of order 1.  For bosonic atoms, this criterion yields the temperature $T_c$ for the onset of Bose-Einstein condensation.  For $T<T_c$, a macroscopic number of particles $N_0(T)$ occupies the lowest single-particle state, and the fraction of particles in this lowest state increases as $T$ decreases.  In an ideal gas, all particles occupy the single-particle ground state at $T=0$ K.  Dilute trapped quantum gases have low densities with $n\sim 10^{13}$ $\rm cm^{-3}$, roughly $10^{-6}$ smaller than room-temperature air.  This density gives a low transition temperature $T_c \sim 10^{-6}$~K.

Alternatively, take a condensed-matter view.  Each particle occupies a ``box'' of dimension $n^{-1/3}$, which gives a zero-point confinement energy $\sim \hbar^2 n^{2/3}/M$.  In the classical limit, the thermal energy $k_BT$ is much larger than the confinement energy, but as $T$ falls, the system eventually reaches the transition temperature $k_BT_c \sim \hbar^2 n^{2/3}/M$ for the onset of quantum degeneracy. For a Bose system, this analysis gives the same criterion for $T_c$ as found from the phase-space density.  For a Fermi system, it gives the usual Fermi temperature $T_F$.  Electrons in metals have a large $T_F\sim 10^4$ K, whereas liquid $^3$He has $T_F\sim 1$ K because the number density remains similar but the mass is larger by roughly $10^4$.  Typical dilute trapped Fermi gases have low number density and large mass, which leads to $T_F \sim 10^{-6}$ K, like a dilute trapped Bose gas.

These ideas apply directly to an ideal Bose gas in a spherical harmonic trap, with trap potential $V_{\rm tr}(r) = \frac{1}{2} M\omega^2 r^2$. The familiar ground-state wave function is a Gaussian $\psi_0(r)\propto \exp(-r^2/2d^2)$, with the characteristic size given by the oscillator length $d=\sqrt{\hbar/(M\omega)}$.  For dilute atomic gases, $d$ is  typically a few $\mu$m.  In a harmonic trap with $N$ particles, the onset of BEC occurs at $k_B T_c \sim \hbar \omega N^{1/3}$; for the typical value $N\sim 10^6$, this yields $T_c\sim 1\  \mu$K.  Above $T_c$ there is only a wide thermal cloud, but below $T_c$ a narrow condensate of width $d$ starts to appear,  rising from the much wider thermal cloud.  As $T\to 0$ K, the thermal cloud disappears, leaving only the narrow condensate.  The presence of a Bose-Einstein condensate at $T=0$ K means that all the $N$ condensed particles form a coherent quantum state described by a macroscopic wave function $\Psi(r) = \sqrt{N} \psi_0(r)$.   The normalization is $\int dV\,|\Psi|^2 = N$ for $T \ll T_c$.  In this ideal gas, the condensate density (which is the same as the total density at low temperature) is nonuniform with $n(r) =|\Psi(r)|^2$.  In a spherical harmonic trap, the condensate density has the Gaussian form $n(r) \propto N\exp(-r^2/d^2)$.

What is the effect of interparticle interactions?  The basic idea was due to Bogoliubov:  for weak interparticle potentials and $T\ll T_c$, nearly all the particles remain in the condensate.  For cold dilute gases, the typical interparticle spacing is $n^{-1/3} \sim $ a few 100 nm.  In contrast, the interactions are short range, characterized by  the $s$-wave scattering length $a\sim$ a few nm.  Thus the dimensionless parameter $na^3 $ is small (of order $10^{-6}$ for a typical cold dilute gas).  Each particle in the condensate experiences a Hartree mean-field potential $V_{\rm H} (\bm r) = (4\pi a \hbar^2/M)\, n(\bm r)= (4\pi a \hbar^2/M) |\Psi|^2$ from the contact interaction with all the remaining condensed particles.  The self-consistent condensate wave function obeys a nonlinear Schr{\"o}dinger equation, usually called the Gross-Pitaevskii (GP) equation
\begin{equation}\label{GP}
\left(-\frac{\hbar^2 \nabla^2}{2M} +V_{\rm tr} + \frac{4\pi a \hbar^2}{M} |\Psi|^2\right) \Psi = \mu \Psi,
\end{equation}
where $\mu$ is the chemical potential.  

In the presence of a trap, this GP equation involves a {\em new} dimensionless ``interaction''  parameter $Na/d$ that  combines the effects of the trap and the interactions.  Note that the typical ratio $a/d$ is small,  of order $10^{-3}$, but with $N \approx 10^6$, this GP interaction parameter is large.  Hence the repulsive interactions expand the condensate to a radius $R$ that significantly exceeds the ideal radius $d$  (a typical value is $R/d \sim 10$).  Consequently, the radial gradient of $\Psi$ becomes small when $Na/d \gg 1$.  In this limit,  one can neglect the kinetic energy term in the GP equation (\ref{GP}),~\cite{Baym96} leading to the much simpler algebraic relation 
\begin{equation}\label{TF}
\frac{4\pi a \hbar^2}{M} |\Psi(\bm r)|^2 = \mu -V_{\rm tr}(\bm r),
\end{equation}
which is called the Thomas-Fermi (TF) approximation.
For a spherical harmonic trap, this result yields the simple particle density (an inverted parabola)
\begin{equation}\label{sph}
n(r)= n(0)\left(1 - \frac{r^2}{R^2}\right),
\end{equation}
where $n(0)= M\mu/(4\pi a \hbar^2)$ is the central density and $R^2 =2\mu/(M\omega^2)$ is the squared condensate radius ($R$ is the classical turning point for a particle with energy $\mu$).

Section 2 reviews the time-dependent Gross-Pitaevskii equation, including the structure of a single vortex in an unbounded condensate.  The more complicated situation of a single vortex in a large trapped condensate is analyzed in Sec.~3, and Sec.~4 summarizes some of the experiments on creation and detection of vortices, including large arrays.  Turbulent vortex systems in trapped condensates are considered in Sec.~5,  where at present only a few experiments exist.  Sections 6 and 7 focus on vortex arrays in the mean-field regime, with two distinct cases:  the Thomas-Fermi regime when the vortex cores are well separated and the density variation is negligible, and the lowest Landau level regime, when the vortex cores overlap and the spatial variation of the density becomes important.   In principle, the system should make a quantum-phase transition to a highly correlated nonsuperfluid state for sufficiently fast rotation (Sec.~8), but experiments have not yet achieved this limit.

\section{Time-dependent Gross-Pitaevskii equation}

This central nonlinear equation was proposed independently by Gross and Pitaevskii in 1961.~\cite{Gros61,Pita61}  It provides an accurate description of low-temperature trapped Bose-Einstein condensates,~\cite{Dalf99,Peth08} in particular 
\begin{enumerate}
\item properties of the ground state,
\item free expansion of the condensate after the confining trap is turned off,
\item collapse for attractive interactions,
\item frequency of low-lying collective modes (at the 1\% level of accuracy).
\end{enumerate}
It comes in two different but equivalent versions.

\subsection{Nonlinear Schr{\"o}dinger equation}
This view emphasizes the quantum aspects of the problem
\begin{equation}\label{TDGP}
i\hbar\frac{\partial \Psi}{\partial t} = \left(-\frac{\hbar^2 \nabla^2}{2M} +V_{\rm tr} + \frac{4\pi a \hbar^2}{M} |\Psi|^2\right) \Psi.
\end{equation}
It can be considered to arise from an energy functional 
\begin{equation}\label{TFenergy}
E[\Psi] = \int dV\left(\frac{\hbar^2 |\nabla\Psi|^2}{2m} + V_{\rm tr}|\Psi|^2 + \frac{2\pi a\hbar^2}{M}|\Psi|^4\right).
\end{equation}
The equilibrium state minimizes this energy functional with fixed normalization $\int dV\,|\Psi|^2 = N$, yielding Eq.~(\ref{GP}) as the resulting Euler-Lagrange equation. In this approach, the chemical $\mu$ in Eq.~(\ref{GP}) serves as a Lagrange multiplier that enforces the constraint of fixed $N$.

\subsection{Hydrodynamic description}
Write the condensate wave function as $\Psi  = \exp(iS)|\Psi|$, which yields hydrodynamic variables:
\begin{enumerate}
\item particle density $n = |\Psi|^2$,
\item velocity $\bm v = \hbar\bm\nabla S/M$, which is irrotational except for singularities,
\item particle current density $\bm j = n\bm v$.
\end{enumerate}
The circulation is defined as $\kappa =\oint_C d\bm l\cdot \bm v$ on a closed path $C$.  As in superfluid $^4$He, the single-valued condensate wave function implies quantized circulation with $\kappa =\  {\rm integer}\times 2\pi \hbar/M$ (note a similar result holds for type-II superconductors, where the electronic charge plays a crucial role).

Substitute the form $\Psi = \exp(iS) |\Psi|$ into the time-dependent GP equation (\ref{TDGP}).  The imaginary part gives the usual conservation of particles $\partial n/\partial t+ \bm \nabla\cdot (n\bm v) = 0$.  The real part gives   a generalized Bernoulli equation that incorporates all the physics of compressible irrotational isentropic hydrodynamics, including quantized vortices and their dynamics.  The one new feature is that the trap potential $V_{\rm tr}$ makes an additional contribution to the vortex motion.

\subsection{Straight singly quantized vortex in bulk fluid}

Gross~\cite{Gros61} and Pitaevskii~\cite{Pita61} independently studied the structure of a singly quantized vortex in an unbounded condensate with bulk density $n$.  This weakly interacting dilute gas provided a toy model for a vortex in dense superfluid $^4$He.  Assume a condensate wave function $\Psi(\bm r) = \sqrt n \exp(i\phi) f(r)$, where $\bm  r$ is in the $xy$ plane with $\phi$ the polar angle.  This choice yields a  flow with circular streamlines
\begin{equation}\label{v}
\bm v(\bm r) = \frac{\hbar}{Mr} \hat{\bm \phi}.
\end{equation}
Note that $\bm v$  diverges as $r\to 0$.  The circulation is $\kappa = 2\pi \hbar/M$ with singular vorticity at the origin $\bm \nabla\times \bm v=  \kappa \hat{\bm z} \delta^{(2)}(\bm r)$.  The chemical potential is $\mu = 4\pi a\hbar^2 n/M$ which is the Hartree energy for a uniform fluid.  The speed of sound is simply $s = \sqrt{\mu/M}$, and real $s$ requires a repulsive interaction with $a > 0$.  The centrifugal barrier forces the radial function $f(r)$ to vanish at the origin with a core radius $\approx$ the healing length $\xi$  fixed by the balance between the kinetic energy and the interaction energy $\mu$
\begin{equation}\label{xi}
\xi = \frac{\hbar}{\sqrt{2M\mu}}.
\end{equation}
The vortex core is large compared to the interparticle spacing for a dilute gas with $na^3\ll 1$.  The circulating flow (\ref{v}) around the vortex becomes supersonic near the core;  from  this perspective,  the vortex core arises from acoustic  cavitation.

Note that Gross and Pitaevskii sought to model a vortex in superfluid $^4$He, which has only a single superfluid component.  For dilute trapped alkalai-metal gases like $^7$Li,  $^{23}$Na  and $^{87}$Rb, however, it is easy to make two-component mixtures with various hyperfine states arising from the spin of the single unpaired valence electron and the nuclear spin.  In this case, the vortices can have more complicated internal structures, along with arrays containing vortices in each component.  Section IV.B.5 of Ref.~\cite{Fett09} briefly summarizes the current situation, in particular the  dynamical experiments of Schweikhard {\it et al.}~\cite{Schw04a}  For simplicity, this article will focus on single-component vortices, although  the first experimental creation of a vortex in a trapped condensate in fact relied on the coupling between two hyperfine states in $^{87}$Rb (see Fig.~2 below).

\section{Single Vortex in a Large Trapped Condensate}

Assume an axisymmetric harmonic trap potential with 
\begin{equation}\label{Vtr}
V_{\rm tr}(\bm r) = V_{\rm tr}(r,z) =\textstyle{\frac{1}{2}} M\left(\omega_\perp^2 r^2 + \omega_z^2 z^2\right).
\end{equation}
For $\omega_z\gg \omega_\perp$, the condensate has a flattened disk shape because of the strong axial confinement, whereas for 
$\omega_z\ll \omega_\perp$, the condensate has an elongated cigar shape because of the strong radial confinement.

The usual experimental procedure is to turn off the trap potential and then take an image of the expanded condensate at a later time.  For definiteness, consider a cigar shape with $\omega_\perp \gg \omega_z$.  Before the expansion, the tight radial confinement has a large radial potential energy and a small axial potential energy.  After the trap is turned off, the condensate expands rapidly in the radial direction and soon acquires a flattened disk shape.  Similarly, an initial disk-shaped condensate expands axially to become elongated.  This behavior is known as the ``reversal of aspect ratio;'' it provided crucial evidence for the existence of a BEC in early experiments (see~\cite{Dalf99,Peth08}).  

Assume that such a trap rotates around $\bm{\hat z}$ with angular speed $\Omega$.  In the rotating frame, the original Hamiltonian changes\cite{Landau} from $H$ to $H' = H- \Omega L_z$.  For equilibrium in this rotating frame, the condensate wave function $\Psi$ minimizes the GP energy functional in the rotating frame
\begin{equation}\label{E'}
E'[\Psi,\Omega] = E[\Psi] - \int dV\,\Psi^*\left(\bm \Omega\cdot \bm r\times \bm p\right) \Psi.
\end{equation}
For simplicity, assume  a rotating disk-shaped condensate of radius $R_\perp$.  Use $E' = E-\Omega L_z$  to determine the energy $E'_0$ of a rotating condensate with no vortex and the energy $E'_1$ of a rotating  condensate with a straight off-center vortex at a distance  $r_0$  from the center.  Define the creation energy for the vortex as $\Delta E' = E'_1-E'_0$.   For a  typical dilute gas, the core radius $\xi$ is a few $\times 10^{-7}$ m.  Detailed analysis~\cite{Svid00} yields the following picture (Fig.~1) of the vortex energy $\Delta E'$  as a function of the radial displacement $r_0$ for various fixed values of $\Omega$.  

  \begin{figure}[h] 
  \includegraphics[width=3in]{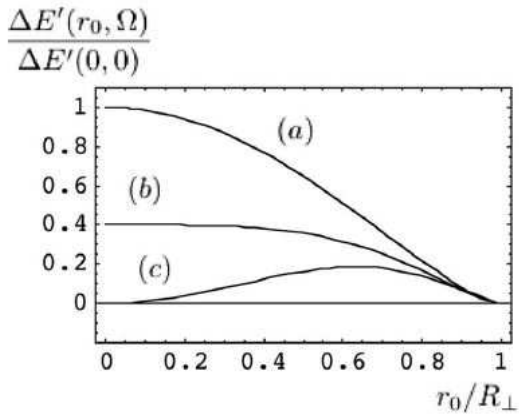}
  \caption{Energy of a single off-center vortex in a disk-shaped condensate for various fixed $\Omega$:  (a) $\Omega = 0$, (b) $\Omega = \Omega_m$ for onset of metastability, (c) $\Omega = \Omega_c$ for onset of thermodynamic stability~\cite{Fett09}  (reprinted with permission of the author and the American Physical Society).}
 \end{figure}

Curve (a) is for $\Omega= 0$, when the energy $\Delta E'$ decreases monotonically with increasing $r_0$, and the trap center is a local maximum of this curve.  In the absence of dissipation, a fixed energy means a fixed radial position $r_0$.  Hence the only allowed motion for  such a vortex is uniform circular precession.  The rate $\dot \phi$  of precession is proportional to the slope of the energy curve at $r_0$   with $\dot\phi \propto -\partial \Delta E'/\partial r_0$.   In the presence of weak dissipation, the vortex moves down the energy curve to reduce its energy, so that the vortex slowly spirals out of the nonrotating condensate.

With  increasing external $\Omega$, the negative central curvature of the energy curve in Fig.~1 decreases, and curve (b) is the special value 
\begin{equation}\label{Omegam}
\Omega_m = \frac{3}{2} \frac{\hbar }{MR_\perp^2}\ln \left(\frac{R_\perp}{\xi}\right)
\end{equation}
when the central curvature vanishes.  This result means that  a central vortex first becomes metastable at $\Omega_m$, because  the trap center is a local {\it minimum} of the energy for $\Omega>\Omega_m$.  For weak dissipation, a vortex would now spiral inward for small lateral displacements to reduce its energy.  Note that such a vortex is not globally stable since $\Delta E'$ remains positive at the origin.  Curve (c) occurs for $\Omega_c = \frac{5}{3} \Omega_m$, when the vortex first becomes truly stable.

How fast does an off-center vortex precess?  One convenient approach relies on the Lagrangian variational approach,~\cite{Lund00,McGee01} where the Lagrangian functional $\cal L$ has the detailed form
\begin{equation}\label{Lagrangian}
{\cal L}[\Psi]  = \frac{i\hbar}{2}\int dV\left(\Psi^*\frac{\partial \Psi }{\partial t} - \frac{\partial \Psi^* }{\partial t} \Psi\right) -E'[\Psi].
\end{equation}
This functional is stationary for small variations of $\Psi$ and $\Psi^*$, with the familiar time-dependent Gross-Pitaevskii equation as the Euler-Lagrange equation.   If the trial wave function contains one or more parameters, the resulting $\cal L$ contains  the time derivative of the parameters because of the explicit appearance of $\partial \Psi/\partial t$ and $\partial \Psi^*/\partial t$.  The resulting Lagrangian serves to study the dynamical evolution of the parameters.

In a disk-shaped TF condensate with radius $R_\perp$, the vortex position $\bm r_0(t)$ provides a simple example of such a time-dependent parameter.  For a nonrotating condensate, the precession rate becomes 
\begin{equation}\label{TFprecess}
\dot\phi = \frac{\Omega_m}{1-r_0^2/R_\perp^2},
\end{equation}
where $\Omega_m = \frac{3}{2} (\hbar/MR_\perp^2)\ln(R_\perp/\xi)$ is the frequency for the onset of metastability given in Eq.~(\ref{Omegam}).  It is notable that the precession is {\it positive}, in the same sense as the fluid flow around the core.  Here, the factor $1-r_0^2/R_\perp^2$ comes from the parabolic radial TF density profile (ultimately this dependence reflects the quadratic harmonic trap).  As discussed below, experiments on trapped BECs confirm this result in considerable detail.  It is instructive to compare this result with that for a similar vortex in incompressible fluid bounded by a rigid cylinder of radius $R_\perp$ (as a model for superfluid $^4$He).  Classical hydrodynamics yields a formally similar result
\begin{equation}\label{CLprecess}
\dot\phi_{\rm cl} = \frac{\hbar}{MR_\perp^2}\frac{1}{1-r_0^2/R_\perp^2},
\end{equation}
but here the denominator arises from the image vortex located at a distance $R_\perp^2/r_0$, instead of the nonuniform density.  Also, there is no ``large'' logarithmic factor $\ln(R_\perp/\xi)$.  

\section{Experimental creation and detection of vortices in BECs}

Most such experiments study equilibrium vortex arrays.  The first vortex was made at JILA (Boulder, CO) in 1999.~\cite{Matt99}  They used a nearly spherical $^{87}$Rb condensate containing two different hyperfine states.  A coherent laser coupling controlled the interconversion between the two species, and a stirring perturbation could spin up the condensate.  When the laser coupling was turned off, they obtained one component with a singly quantized vortex that circulated around a nonrotating core of the other component.  Selective laser tuning provided nondestructive images of either component.  These  images (Fig.~2) allowed a study of the precession of such a two-component vortex around the trap center.~\cite{Ande00}  The core can be as large as 5-10 $\mu$m, depending on the fraction of the nonrotating component, which is readily imaged with visible light.

 \begin{figure}[h]
  \includegraphics[width=4.5in]{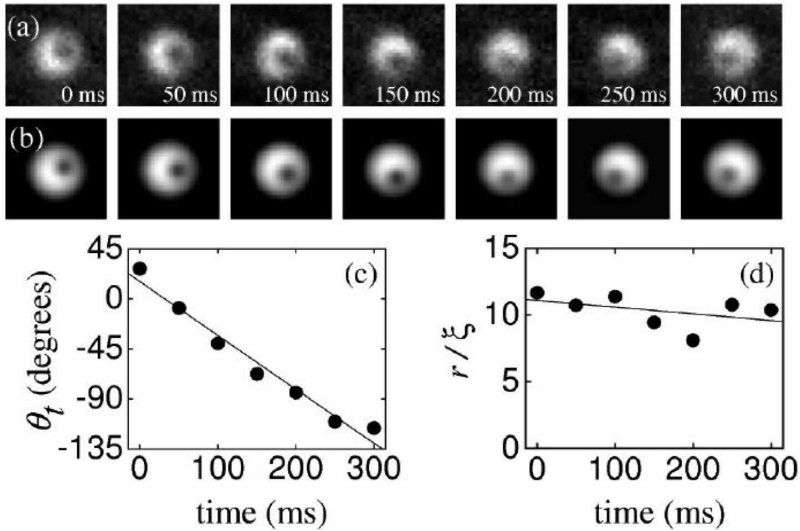}
  \caption{Precessing two-component vortex (a) direct images, at 50 ms intervals (b) smoothed images, (c) angular position, and (d) gradual shrinkage of core due to spin-flip transitions~\cite{Ande00}  (reprinted with permission of the authors and the American Physical Society).}
 \end{figure}
 
The JILA group could also remove the nonrotating core component with an intense laser pulse, leaving a single component vortex with an empty core, which marked  the initial position.    They waited a variable time and then turned off the trap, allowing a visualization of the final position of the vortex.  The observed precession rate agreed with the theoretical analysis at the $\pm 10\%$ level.  The experiments saw no outward radial motion for $\sim 1$ s, implying that dissipation is small on this time scale.

The Ecole Normale Sup{\'e}rieure (ENS) group in Paris studied vortex creation in a very elongated rotating cigar-shaped condensate with one component.   They used an off-center toggled rotating laser beam to deform the transverse trap potential and stir the condensate at an applied frequency  $\Omega/2\pi\lesssim 200$ Hz.  They observed small arrays of up to 11 vortices arranged in two concentric circles (Fig.~3). They needed to expand the condensate to obtain these pictures.~\cite{Dali00}  Note that these images are like patterns predicted and seen in superfluid $^4$He (Fig.~4).~\cite{Yarm79}

\begin{figure}[t]
\includegraphics[width=2.5in]{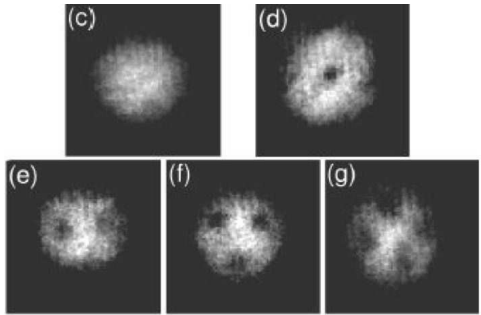}
\caption{Images of small vortex clusters in a rotating BEC~\cite{Dali00}  (reprinted with permission of the authors and the American Physical Society).}
\end{figure}

\begin{figure}[h]
\includegraphics[width=1.75in]{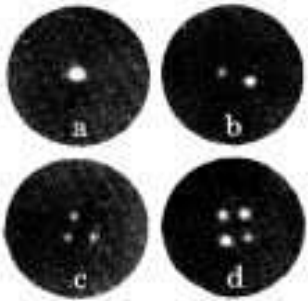}
\caption{Images of small vortex clusters in rotating superfluid $^4$He~\cite{Yarm79}  (reprinted with permission of the authors and the American Physical Society).}
\end{figure}

Soon afterward, the MIT group prepared considerably larger rotating condensates in a less elongated trap.~\cite{Abo01}  They observed large triangular arrays with up to 130 vortices, like the Abrikosov vortices of quantized flux lines in type-II superconductors.  As an alternative approach, Cornell's group at JILA started from a rapidly rotating normal cloud and then cooled into the superfluid state that contained  a   vortex array to accommodate the large angular momentum (Fig.~5).~\cite{Halj01,Enge03}
~ \begin{figure}[h]
\includegraphics[width=1in]{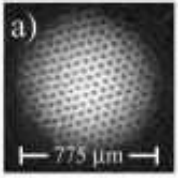}
\caption{Large triangular vortex array in a rotating BEC~\cite{Enge03}  (reprinted with permission of the authors and the American Physical Society).}
\end{figure}
 \begin{figure}[h]
\includegraphics[width=5in]{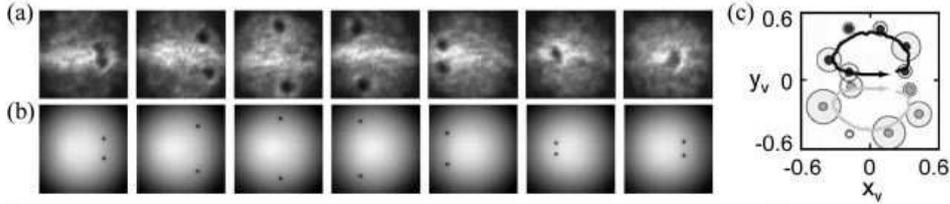}
\caption{Creation of vortex dipole by a blue-detuned laser beam in a disk-shaped TF condensate.  (a) Experimental data, (b) numerical simulation, (c) comparison of observed and theoretical trajectory~\cite{Neel09}  (reprinted with permission of the authors and the American Physical Society).}
\end{figure}

Anderson's group in Arizona has created vortex dipoles ($\pm$ vortex pairs) in a disk-shaped TF condensate.~\cite{Neel09}
They used an intense blue-detuned laser beam as an obstacle in the condensate, moving the condensate so that it sweeps smoothly past the obstacle at differing rates.  Above a critical rate, they created a vortex dipole at a reproducible position, waited a variable time, and then turned off the trap.  Figure 6 shows  expanded pictures of the dynamical motion of the vortex dipole at intervals of 200 ms.  The first row is experimental data, and the second row is a theoretical simulation.  The figure (c) on the right compares the measured trajectory (dots) with the theoretical trajectory (continuous line)

\section{Turbulent vortex systems}

Kobayashi and Tsubota~\cite{Koba07} have carried out numerical simulations of the time-dependent GP equation with  rotations first  about the  $\bm {\hat z}$ axis and then about   the $\bm{\hat x}$ axis, leading to an effectively time-dependent rotation axis $\bm\Omega(t)$ because of the combined rotations.  For a slightly asymmetric triaxial condensate, they find that the condensate surface eventually becomes irregular, and a turbulent vortex tangle then develops for sufficiently long times.
% \begin{figure}[h]
%\includegraphics[width=1.8in]{kobayashi1.eps}
%\caption{Schematic illustration of induced response to simultaneous rotations about two perpendicular axes~\cite{Koba07}  (reprinted with permission of the authors and the American Physical Society).}
%\end{figure}

Bagnato's group in S{\~a}o Carlos, Brazil use a related scheme of oscillations about two axes  to generate what appears to be a turbulent vortex tangle in a cigar-shaped condensate.~\cite{Henn09}  When the trap is turned off, the ``turbulent'' condensate expands with an approximately self-similar profile, in contrast to the usual reversal of aspect ratio for a nonrotating condensate (Fig.~7).  For a uniform vortex array, the vorticity induces an additional  expansion in the two  perpendicular directions.  Thus a combination of random turbulent vorticity in all three directions may explain this unexpected  self-similar behavior.

 \begin{figure}[h]
\includegraphics[width=2.5in]{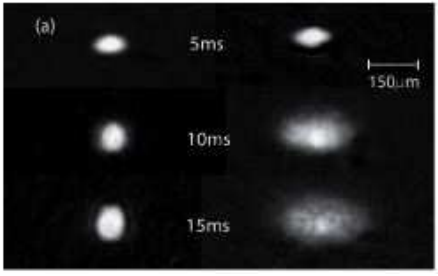}
\caption{Comparison of expansion of nonrotating condensate (left side) and ``turbulent'' condensate (right side)~\cite{Henn09}  (reprinted with permission of the authors and the American Physical Society).}
\end{figure}
\section{Vortex arrays in mean-field Thomas-Fermi regime}

As expected, the mean vortex density $n_v$ in rotating condensates obeys the Feynman relation familiar from superfluid $^4$He, with $n_v = 2\Omega/\kappa = M\Omega/(\pi \hbar)$.  Correspondingly, the area per vortex is $1/n_v = \pi\hbar/(M\Omega)\equiv \pi l^2$, which defines the radius $l = \sqrt{\hbar/(M\Omega)}$ of an equivalent circular cell.  Note that the intervortex spacing $\sim 2l$ decreases like $1/\sqrt\Omega$.

With increasing  $\Omega $, the mean vortex density grows linearly following the Feynman relation.  In addition,  centrifugal forces expand the condensate radially, so that the area $\pi R_\perp^2 $ also increases.  Hence the number of vortices $N_v = M\Omega R_\perp^2 /\hbar $ increases faster than linearly with $\Omega$.  Conservation of particles  implies that the condensate also shrinks axially.  The Thomas-Fermi approximation assumes that the interaction energy $\langle \frac{1}{2}g|\Psi|^4\rangle$ and the trap energy $\langle V_{\rm tr}|\Psi|^2\rangle$ are both  large relative to the gradient energy for density variations $(\hbar^2/2M)\langle (\bm\nabla |\Psi|)^2\rangle$.  This TF approximation holds for well-separated vortices with $l\gg \xi$, but it breaks down when the vortex lattice becomes ``dense'' and the cores start to overlap.

For a quantitative description,  note that the kinetic energy involves $-i\bm\nabla \Psi\approx M\bm v\Psi/\hbar$ from the gradient of the phase, since the density variation is here negligible.  Hence the TF energy functional in the rotating frame in Eq.~(\ref{E'}) becomes 
\begin{equation}\label{E'TF}
E'[\Psi] = \int dV \left[\left(\textstyle{\frac{1}{2}}Mv^2 +V_{\rm tr} -M\bm \Omega \cdot \bm r\times \bm v\right)|\Psi|^2 +\textstyle{\frac{1}{2}} g |\Psi|^4\right],
\end{equation}
where $g= 4\pi a\hbar^2/M$ is the coupling constant and $\bm v$ is the velocity generated by all the vortices.  In the present limit of many vortices, the Feynman relation implies that this velocity is just the solid-body result $\bm v_{\rm sb}=\bm \Omega\times \bm r$.   For $\Omega$ along $\hat{\bm z}$, substitution into Eq.~(\ref{E'TF}) yields 
\begin{equation}\label{E'TFa}
E'[\Psi] = \int dV \left[\textstyle{\frac{1}{2}}M\left(\omega_\perp^2-\Omega^2\right)|\Psi|^2+ \textstyle{\frac{1}{2}}M\omega_z^2 |\Psi|^2+\textstyle{\frac{1}{2}} g |\Psi|^4\right],
\end{equation}
which now looks exactly like the TF energy for a nonrotating condensate, but with a {\it reduced} squared radial  trap frequency 
$\omega_\perp^2 \to \omega_\perp^2 - \Omega^2$.

Hence the TF condensate density  now depends explicitly on $\Omega$: 
$ |\Psi(r,z)|^2 = n(0)\left(1-r^2/R_\perp^2 -z^2/R_z^2\right)$, with 
\begin{equation}\label{rad}
R_\perp^2 = \frac{2\mu}{M(\omega_\perp^2 -\Omega^2)} \quad\hbox{and}\quad  R_z^2 = \frac{2\mu}{M\omega_z^2}.
\end{equation}
It is clear that $\Omega$ cannot exceed $\omega_\perp$, since otherwise the radial confinement would disappear.  In addition, the central density $n(0)$ and the chemical potential $\mu = gn(0)$ both  decrease with increasing $\Omega$ because of the reduced radial confinement.  The formulas (\ref{rad})  for the condensate radii show how the aspect ratio changes with $\Omega$
\begin{equation}\label{aspect}
\frac{R_z(\Omega)}{R_\perp(\Omega)} = \frac{\sqrt{\omega_\perp^2-\Omega^2}}{\omega_z}.
\end{equation}
This last effect provides an important diagnostic tool to determine the actual angular velocity $\Omega$ (Fig.~8).~\cite{Schw04}  The measured aspect ratio indicates that $\Omega/\omega_\perp$ can become as large as $\approx 0.993$.
\begin{figure}[h]
 \includegraphics[width=2.5in]{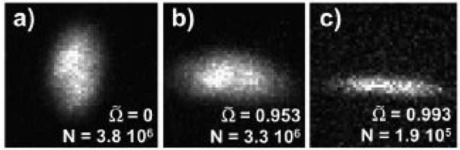}
 \caption{Increasing angular velocity dramatically flattens an initially cigar-shaped condensate~\cite{Schw04}  (reprinted with permission of the authors and the American Physical Society).}
\end{figure}

In 1966, Tkachenko~\cite{Tkac66} studied the collective modes of an infinite triangular vortex  lattice for motion perpendicular to the vortex axes.  He predicted a special mode involving long-wavelength transverse shearing motion of the straight vortex lines.  This behavior arises from the discrete quantized vorticity in each vortex.  It disappears for ``inertial waves'' in a rotating classical fluid with uniform vorticity (namely, $\kappa\to 0$ such that $n_v\kappa= 2\Omega$).

Cornell's group at  JILA studied the dynamical motion for such Tkachenko waves in trapped BECs.  They formed a uniform vortex array and then applied a weak perturbation,~\cite{Codd03} setting up these Tkachenko waves.  Figure 9 shows the deformed vortex lattice at $\frac{1}{4}$ and $\frac{3}{4}$ of the oscillation period, with clear evidence of the phase reversal.  The observed motion has the correct quantitative form, but the measured period differs from the predictions.  Sonin's review article~\cite{Soni87}  discusses earlier studies of Tkachenko waves in superfluid $^4$He (see Sec.~VI.E, particularly Fig.~3).
\begin{figure}[ht]
 \includegraphics[width=2in]{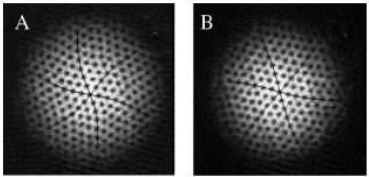}
 \caption{Deformed vortex lattice at 1/4 and 3/4 period, showing the transverse Tkachenko wave~\cite{Codd03} (reprinted with permission of the authors and the American Physical Society).}
\end{figure}

\section{Vortex arrays in  mean-field lowest Landau level regime}

When the vortex cores begin to overlap, it becomes necessary to include the kinetic energy arising from the density variation near each vortex core.  Evidently, the Thomas-Fermi approximation breaks down, for it ignores such rapid density variations.  Hence it is essential to return to the full GP energy functional [Eq.~(\ref{E'})] for $E'[\Psi,\Omega]$ in the rotating frame.  In this rapid-rotation limit ($\Omega\lesssim\omega_\perp$), Ho~\cite{Ho01} pointed out that it is possible to incorporate the full kinetic energy exactly.  The centrifugal forces expand the condensate, which becomes disk-shaped.  For simplicity, it is convenient to treat a two-dimensional circular condensate that is uniform in the $z$ direction over a length $Z$.  The full condensate wave function $\Psi(\bm r,z)$ can then be written as $\sqrt{N/Z} \,\psi(\bm r)$, where $\psi(\bm r)$ is a two dimensional wave function with unit normalization $\int d^2 r\,|\psi|^2 = 1$.

The general two-dimensional GP energy functional in the rotating frame becomes 
\begin{equation}\label{E'LLL}
 E'[\psi,\Omega] =\int d^2r\,\psi^*\left(\underbrace{\frac{p^2}{2M} + \frac{1}{2}M\omega_\perp^2r^2 -\Omega L_z}_{\rm one-body\  oscillator\  {\cal H}_0'}+ \underbrace{\frac{1}{2} g_{\rm 2D}|\psi|^2}_{\rm interaction}\right)\psi,
\end{equation}
where $\bm p = -i\hbar\bm \nabla$, $L_z=\hat{\bm z}\cdot\bm r\times \bm p$, and $g_{\rm 2D} = Ng/Z$.   The one-body oscillator hamiltonian ${\cal H}_0'$ in the rotating frame is exactly soluble and has the following eigenvalues~\cite{Cohe77}
\begin{equation}\label{eigen}
\epsilon_{nm} = \hbar[\omega_\perp+n(\omega_\perp+\Omega) + m(\omega_\perp-\Omega)],
\end{equation}
where $n$ and $m$ are non-negative integers.  In the limit $\Omega\to \omega_\perp$, these eigenvalues are essentially independent of $m$, which implies a large degeneracy.  The other quantum number $n$ then becomes the Landau-level index. The lowest Landau level with $n= 0$ is separated from the higher Landau levels by a gap $\sim 2\hbar \omega_\perp$.

The large radial expansion means a small central density $n(0)$, so that the interaction energy $g_{\rm 2D}n(0)$ eventually becomes small compared to the gap $2\hbar\omega_\perp$.  In this limit, it is natural to focus on the lowest Landau level (LLL), with $n=0$ and non-negative $m\ge 0$.  The ground-state wave function is a Gaussian $\psi_{00}\propto \exp(-r^2/2d_\perp^2)$, where $d_\perp=\sqrt{\hbar/M\omega_\perp}$ is analogous to the magnetic length in the original Landau problem of an electron in a uniform magnetic field.  The general LLL  eigenfunctions have a very simple form
 \begin{equation}\label{LLL}
\psi_{0m}(\bm r) \propto r^m e^{im\phi} e^{-r^2/2d_\perp^2}.
\end{equation}
Since $r\exp(i\phi)$ is just the polar form of the complex variable $\zeta = x+iy$, these LLL eigenfunctions become $\psi_{0m}\propto \zeta^m \exp(-r^2/2d_\perp^2)$ with $m\ge0$.  Apart from the ground-state Gaussian, this is just $\zeta^m$, a non-negative power of the complex variable $\zeta$.

Assume that the GP condensate wave function is a {\it finite} linear combination of these LLL eigenfunctions
\begin{equation}\label{LLLtrial}
\psi(\bm r) _{\rm LLL} = \sum_{m\ge 0} c_m\psi_{0m}(\bm r) \equiv f(\zeta) e^{-r^2/2d_\perp^2},
\end{equation}
where $f(\zeta)= \sum_{m\ge 0} c_m\zeta^m$ is an {\it analytic function} of the complex variable $\zeta$.  Specifically, $f(\zeta)$ is a complex polynomial and can be factorized as $f(\zeta) = \prod_j\left(\zeta-\zeta_j\right)$ apart from an overall constant.  Note that $f(\zeta)$ vanishes at each of the points $\{\zeta_j\}$, which are the positions of the nodes of of $\psi_{\rm LLL}$.  In addition, the phase of the wave function increases by $2\pi$ whenever $\zeta$ moves around any of these zeros $\{\zeta_j\}$ in the positive sense.  As a result, the LLL trial function (\ref{LLLtrial}) has singly quantized vortices located at the positions of the zeros $\{\zeta_j\}$.

The mean-field LLL regime implies the striking result  that the spatial distribution of the vortices completely determines the spatial variation of the number density $n(\bm r) = |\psi_{\rm LLL}(\bm r)|^2$.  The core size is comparable with the intervortex spacing $l = \sqrt{\hbar/M\Omega}$,~\cite{Fisc03} which is the same as $d_\perp$ in the limit $\Omega\approx \omega_\perp$.  Unlike the mean-field TF regime at lower $\Omega$, here the wave function $\psi_{\rm LLL}$ includes all the kinetic energy.  Since the LLL wave functions play a crucial role in the quantum Hall effect (two-dimensional electrons in a strong magnetic field), this LLL regime is sometimes called the ``mean-field quantum Hall'' regime.

It is important to emphasize that we are still in a regime governed by the GP equation, so there is still a BEC with a macroscopic condensate wave function.  The corresponding many-body ground state is simply a Hartree product  with each particle in the {\it same} one-body solution $\psi_{\rm LLL}(\bm r)$:
\begin{equation}\label{PsiGP}
\Psi_{\rm GP}(\bm r_1, \bm r_2,\cdots, \bm r_N)\propto \prod_{j = 1}^N \psi_{\rm LLL}(\bm r_j).
\end{equation}
This is a coherent superfluid state since the GP single-particle state $\psi_{\rm LLL}$ has macroscopic occupation.

\section{Beyond the GP picture: quantum phase transition to highly correlated states}

As $\Omega$ increases still closer toward $\omega_\perp$, the question of what happens  beyond the mean-field LLL regime
 remains a subject of vigorous debate.~\cite{Coop08,Vief08}  Generally, a quantum phase transition is predicted to take place from the  coherent many-body BEC ground state in Eq.~(\ref{PsiGP}) to one of various correlated many-body states that are not superfluid and do not have  macroscopic occupation.

To quantify the discussion, it is conventional to define the ratio
$\nu \equiv N/N_v$, which is the number of atoms per vortex.  Because of similarities to the two-dimensional electron gas in a strong magnetic field, the ratio $\nu$ is called the ``filling fraction.''  Current experiments have $N\sim 10^5$ and $N_v \sim$ a few  hundred vortices so that the typical $\nu\sim$ a few hundred.  

Numerical studies for small number of vortices ($N_v\lesssim 8$)  and variable number of particles $N$ indicate that the GP coherent state is favored for $\nu \gtrsim 6-8$.~\cite{Coop01}  For smaller $\nu$, however, the ground state typically has a very different form.  Specifically, analytical solutions for small $N$ and large angular momentum $L$ are not of the factorized GP form.~\cite{Wilk00}

For $\nu\lesssim6-8$, the ground state is predicted to be one of a sequence of highly correlated states similar to some of those known for the quantum Hall effect for electrons.  One particularly simple example is a bosonic version of the Laughlin state
\begin{equation}\label{PsiL}
\Psi_{\rm Laughlin}(\bm r_1, \bm r_2, \cdots, \bm r_N)\propto \prod_{n<n'}^N \left(z_n-z_{n'}\right)^2\exp\left(-\sum_{n=1}^N\frac{|z_n|^2}{2d_\perp^2}\right),
\end{equation}
where $z_n = x_n + iy_n$ refers to the $n$th particle.  The original Laughlin state for electrons had a power 3 in the double product to ensure antisymmetry, whereas the present power 2 ensures symmetry, as appropriate for bosons.  These correlated many-body states are qualitatively different from the coherent GP form.  Specifically, the double product in (\ref{PsiL}) involves $N(N-1)/2$ factors for all possible pairs and vanishes whenever any two particles are close together.  This last factor is the source of correlations, for it reduces the energy in the typical case of short-range repulsive potentials.

How might one reach the correlated regime?  The essential step is to reduce the ratio $\nu = N/N_v$ (the number of atoms per vortex).  One possibility is to use an elongated condensate with a  relatively large vortex array.  Subsequent  application of an optical lattice along the rotation axis would leave an array of thin vortex-filled disks that might achieve this goal.  

Another possible idea is to use laser-induced synthetic vector potentials that can mimic the effect of rotation.~\cite{Gunt09,Lin09,Spie09,Murr09}  Spielman's group at NIST (Washington DC) has indeed produced vortices with this scheme,~\cite{Spie10} although they do not see regular arrays.  This intriguing approach will certainly receive more attention.

\begin{acknowledgements}
I am grateful to B.\ Anderson, V.\ Bagnato, A.\ Golov, and E.\ Sonin   for  discussions and comments during the preparation of this manuscript.
\end{acknowledgements}

\pagebreak


\begin{thebibliography}{99}

\bibitem{Dalf99}    F.~Dalfovo, S.~Giorgini, L.~P.~Pitaevskii, and S.~Stringari,
                        Rev.~Mod.~Phys.~{\bf 71}, 463 (1999).
\bibitem{Fett01a}  A.~L.~Fetter and A.~A.~Svidzinsky,  J.~Phys.:\ Condens.~Matter {\bf13}, R135 (2001).
\bibitem{Fett02}  A.~L.~Fetter, J.~Low Temp.~Phys.~{\bf 129}, 263 (2002).
\bibitem{Peth08} C.\ J.\ Pethick and H.\ Smith, {\it Bose-Einstein\ Condensation in Dilute Gases} (Cambridge University Press, Cambridge, 2008), second ed.
\bibitem{Fett09} A.\ L.\ Fetter, Rev.\ Mod.\ Phys.\ {\bf 81}, 647 (2009).
\bibitem{Baym96}  G.~Baym and C.~J.~Pethick,  Phys.~Rev.~Lett.~{\bf 76}, 6 (1996).
\bibitem{Gros61}  E.~P.~Gross, Nuovo Cimento {\bf 20}, 454 (1961).
\bibitem{Pita61}  L.~P.~Pitaevskii, Zh.~Eksp.~Teor.~Fiz.~{\bf 40}, 646 (1961) [Sov.~Phys.~JETP {\bf 13}, 451 (1961)].
\bibitem{Schw04a}  V.~Schweikhard,  I.~Coddington, P.~Engels, S.\ Tung, and E.\ A.\ Cornell, Phys.~Rev.~Lett.~{\bf 93}, 210403  (2004).
\bibitem{Landau}    L.~D.~Landau and E.~M.~Lifshitz,
                        {\it Mechanics}, Pergamon Press, Oxford (1960);
                    E.~M.~Lifshitz and L.~P.~Pitaevskii,
                        {\it Statistical Physics}, Pergamon Press, Oxford (1980).
\bibitem{Svid00}    A.\ A.\ Svidzinsky and A.\ L.\ Fetter, 
                        Phys.\ Rev.\ Lett.\ {\bf 84}, 5919 (2000).
\bibitem{Lund00}    E.~Lundh and P.~Ao,
                        Phys.~Rev.~A~{\bf 61}, 063612 (2000).
\bibitem{McGee01}   S.~A.~McGee and M.~J.~Holland, 
                        Phys.~Rev.~A~{\bf 63}, 043608 (2001).
\bibitem{Matt99}    M.\ R.\ Matthews, B.\ P.\ Anderson, P.\ C.\ Haljan, D.\ S.\ Hall, 
                    C.\ E.\ Wieman, and E.\ A.\ Cornell,
                        Phys.\ Rev.\ Lett.\ {\bf 83}, 2498 (1999).
\bibitem{Ande00} B.~P.~Anderson,  P.~C.~Haljan, C.~E.~Wieman, and E.~A.~Cornell, Phys.~Rev.~Lett.~{\bf 85}, 2857 (2000).
\bibitem{Dali00}    K.~W.~Madison, F.~Chevy, W.~Wohllenben, and J.~Dalibard,
                        Phys.~Rev.~Lett.~{\bf 84}, 806 (2000).
%\bibitem{Madi00}    K.~W.~Madison, F.~Chevy, W.~Wohllenben and J.~Dalibard, 
                    %    J.~Mod.~Opt.~{\bf 47}, 2725 (2000).
\bibitem{Yarm79} E.\ J.\ Yarmchuk, M.\ J.\ V.\ Gordon, and R.\ E.\ Packard, Phys.\ Rev.\ Lett.\  {\bf 43}, 214 (1979).
\bibitem{Abo01}     J.~R.~Abo-Shaeer, C.~Raman, J.~M.~Vogels, and W.~Ketterle,
                        Science {\bf 292}, 476 (2001).     
\bibitem{Halj01}   P.\ C.\ Haljan,  I.\ Coddington, P.\ Engels, and E.\ A.\ Cornell, Phys.\ Rev.\ Lett.\  {\bf 87}, 210403 (2001).          
                        
\bibitem{Enge03}  P.\  Engels,  I.\ Coddington, P.\ C.\ Haljan, V.\ Schweikhard, and E.\ A.\ Cornell, Phys.\ Rev.\ Lett.\  {\bf 90}, 170405 (2003).
% \bibitem{ Sche07}  D.\ R.\ Scherer, C.\ N.\ Weiler, T.\ N.\  Neely, and B.\ P.\ Anderson, Phys.\ Rev.\ Lett.\ {\bf 98}, 110402 (2007).                  
\bibitem{Neel09}  T.\ W.\ Neely, E.\ C.\ Samson, A.\ S.\ Bradley, M.\ J.\ Davis, and B.\ P.\ Anderson, Phys.\ Rev.\ Lett.\  {\bf 104}, 160401 (2010).
\bibitem{Koba07}  M.\ Kobayashi and M.\ Tsubota, Phys.\ Rev.\ A {\bf 76}, 045603 (2007).
\bibitem{Henn09}  E.\ A.\ L.\ Henn, J.\ A.\ Seman, G.\ Roati,  K.\ M.\ F.\ Magalh{\~a}es, and V.\ S.\ Bagnato, Phys.\ Rev.\ Lett.\  {\bf 103}, 045301 (2009).
 
\bibitem{Schw04} V.~Schweikhard,  I.~Coddington, P.~Engels, V.\ P.\ Mogendorff, and E.\ A.\ Cornell, Phys.~Rev.~Lett.~{\bf 92}, 040404 (2004).


%\bibitem{Fisc03}  U.\ R.\ Fischer and G.\ Baym, Phys.~Rev.~Lett.~{\bf 90}, 140402 (2003).

\bibitem{Tkac66}    V.~K.~Tkachenko, 
                        Zh.~Eksp.~Teor.~Fiz.~{\bf 49}, 1875 (1965)
                           [Sov.~Phys.~JETP {\bf 22}, 1282 (1966)];
                        Zh.~Eksp.~Teor.~Fiz.~{\bf 50}, 1573 (1966)
                            [Sov.~Phys.~JETP {\bf 23}, 1049 (1966)].
                            
\bibitem{Codd03}  I.~Coddington, P.\  Engels, V.\  Schweikhard, and E.~A.~Cornell,  Phys.\ Rev.\ Lett.\  {\bf 91}, 100402 (2003).

%\bibitem{Fett67}    A.~L.~Fetter,
                       %Phys.~Rev.~{\bf 162}, 143 (1967).
%\bibitem{Fett75}    A.~L.~Fetter,
                     %   Phys.~Rev.~B {\bf 11}, 2049 (1975).

\bibitem{Soni87}    E.~B.~Sonin,  Rev.~Mod.~Phys.~{\bf 59}, 87 (1987).

%\bibitem{Soni04}  E.\  B.\ Sonin, Phys.\ Rev.\ A {\bf 71}, 011603(R) (2005).

%\bibitem{Baym03}  G.\ Baym, Phys.~Rev.~Lett.~{\bf 91}, 110402 (2003).

%\bibitem{Angl02}  J.\ R.\ Anglin and M.\  Crescimanno, cond-mat/0210063.
%
%\bibitem{Codd03}    I.~Coddington, P.~Engels, V.~Schweikhard and E.~A.~Cornell,
                        %Phys.~Rev.~Lett.~{\bf 91}, 100402 (2003).

%\bibitem{Zwie05}  M.\ Zwierlein, J.\ R.\ Abo-Shaeer, A.\  Schirotzek, C.\ H.\ Schunck,  and W.\ Ketterle, Nature {\bf 435}, 1047 (2005).

%\bibitem{Sens06}  R.~Sensarma, M.~Randeria, and T.-L.~Ho, Phys.~Rev.\ Lett.\ {\bf 96},  090403 (2006).
	
\bibitem{Ho01}  T.-L.\ Ho, Phys.\ Rev.\ Lett.\ {\bf 87}, 060403 (2001).

\bibitem{Cohe77}  C.\ Cohen-Tannoudji, B.\ Diu and F.\ Lalo\"e, {\it Quantum Mechanics} (J.\ Wiley \& Sons, New York, 1977), Volume I, pp.\ 742-764.

\bibitem{Fisc03}  U.\ R.\ Fischer and G.\ Baym, Phys.~Rev.~Lett.~{\bf 90}, 140402 (2003).

\bibitem{Coop08} N.\ R.\ Cooper, Adv.\ Phys.\ {\bf 57}, 539 (2008)

\bibitem{Vief08}  S.\ Viefers, J.\ Phys.:\ Condens.\ Matter {\bf 20}, 12302 (2008).

%\bibitem{Wata04}  G.\ Watanabe, G.\ Baym, and C.\ J.\ Pethick, Phys.\ Rev.\ Lett.\ {\bf 93}, 190401 (2004).

%\bibitem{Afta05}  A.\ Aftalion, X.\ Blanc, and J.\ Dalibard, Phys.\ Rev.\ A {\bf 71}, 023611 (2005).

%\bibitem{Coop04}  N.\ R.\ Cooper, S.\ Komineas, and N.\ Read,  Phys.\ Rev.\ A {\bf 70}, 033604 (2004).

%\bibitem{Wilk00}  N.\ K.\  Wilkin and M.\ J.\ F.\  Gunn, Phys.\ Rev.\ Lett.\ {\bf 84}, 6 (2000).

\bibitem{Coop01}  N.\ R.\ Cooper, N.\ K.\  Wilkin, and M.\ J.\ F.\  Gunn, Phys.\ Rev.\ Lett.\ {\bf 87}, 120405 (2001).

\bibitem{Wilk00}  N.\ K.\  Wilkin and M.\ J.\ F.\  Gunn, Phys.\ Rev.\ Lett.\ {\bf 84}, 6 (2000).

%\bibitem{Sino02} J.\ Sinova, C.\ B.\ Hanna, and A.\ H.\ MacDonald, Phys.\ Rev.\ Lett.\  {\bf 89}, 030403 (2002).

%\bibitem{Baym04}  G.\ Baym, Phys.\ Rev.\  A {\bf 69}, 043618 (2004).

%\bibitem{Vala56}  J.\ G.\ Valatin, Proc.\ Roy.\ Soc.\ {\bf 238}, 132 (1956).

%%\bibitem{Fett07}  A. L. Fetter, Phys.\ Rev.\ A {\bf 75}, 013620 (2007). 

%\bibitem{Sinh05}  S.\ Sinha and G.\ V.\  Shlyapnikov, Phys.\  Rev.\ Lett.\ {\bf 94}, 150401 (2005).

%\bibitem{Sanc05}  P.\  S\'anchez-Lotero and J.\  J.\ Palacios, Phys.\ Rev.\ A {\bf 72}, 043613 (2005).

\bibitem{Gunt09}  K.\ J.\ G{\"u}nter, M.\ Cheneau, T.\ Yefsah, S.\ P.\ Rath, and J.\  Dalibard, Phys.\ Rev.\ A {\bf 79}, 011604(R) (2009).
\bibitem{Lin09} Y.-J.\ Lin, R.\ L.\ Compton, A.\ R.\ Perry, W.\ D.\ Phillips, J.\ V.\ Porto, and I.\ B.\ Spielman, Phys.\ Rev.\ Lett.\ {\bf 102}, 130401 (2009).
\bibitem{Spie09}  I.\ B.\ Spielman, Phys.\ Rev.\ A {\bf 79},  063613 (2009).
\bibitem{Murr09}  D.\ R.\ Murray, P.\ {\"O}hberg, D.\ Gomila, and S.\ M.\  Barnett, Phys.\ Rev.\ A {\bf 79},  063618 (2009).
\bibitem{Spie10}  Y.-J.\ Lin, R.\ L. Compton, K.\ Jim{\'e}nez-Garc{\'{\i}}a, J.\ V.\  Porto, and  I.\ B.\ Spielman, Nature {\bf 462}, 628 (2009).



\end{thebibliography}
\end{document}